# Symmetry breaking in hexagonal and cubic polymorphs of $BaTiO_3$

Sina Hashemizadeh, Alberto Biancoli, Dragan Damjanovic[*]

*Ceramics Laboratory, Swiss Federal Institute of Technology in Lausanne – EPFL, 1015 Lausanne, Switzerland*

$BaTiO_3$ appears in cubic and hexagonal variants, both of which are centrosymmetric. Samples of cubic $BaTiO_3$ are known to exhibit breaking of the centric symmetry locally and globally. It has been proposed that the local symmetry breaking originates in polar regions, the precursors of the ferroelectric phase. Origins of the macroscopic symmetry breaking, which are not well understood, have been previously tentatively correlated with inhomogeneities in the samples, such as strain gradients that may align or redistribute objects such as charged point defects or polar regions making material macroscopically polar. No such data are available for $BaTiO_3$ with hexagonal symmetry. We compare dielectric, elastic, and pyroelectric properties of the two materials in polycrystalline form. In contrast to cubic $BaTiO_3$, hexagonal $BaTiO_3$ does not exhibit macroscopic pyroelectric response at room temperature. This is consistent with apparent absence of polar regions in the hexagonal material and the fact that in hexagonal $BaTiO_3$ strain rather then polarization is the order parameter for the phase transition into ferroelectric-ferroelastic phase. The thermally stimulated currents measured in hexagonal and cubic $BaTiO_3$, however, show that both materials exhibit noncentric macroscopic symmetry. This result supports the idea that extrinsic factors such as strain gradients, which are apparently common for both materials, may break the macroscopic symmetry, which may then lead to alignment and redistribution of polar regions or charged defects.

[*] Corresponding author: dragan.dmjanovic@epfl.ch

# I. INTRODUCTION

Physical effects that are limited by symmetry to noncentrsymmetric or polar structures, such as piezoelectricity, pyroelectricity, and optical second harmonic generation, have been reported in materials that nominally exhibit an inversion center.[1-5] Ferroelectric materials are here of a special interest. Below Curie temperature ($T_C$) ferroelectric crystals exhibit reversible polarization, and therefore, a polar symmetry. Above $T_C$, in their paraelectric phase, most ferroelectrics are centrosymmetric. For those ferroelectrics, pyroelectricity and piezoelectricity should be observed only below $T_C$.[6] Nevertheless, the effects that are forbidden by a centric symmetry have been reported in paraelectric phases of several ferroelectric materials.[4,6-12] Origins of the symmetry breaking are multiple and are, in general, not well understood.[6] They have been associated with randomly distributed polar regions (or precursors of the ferroelectric state)[12] that may form in ferroelectrics hundreds of degrees above $T_C$, leading to a local breaking of the centric symmetry, but preserving macroscopic centrosymmetricity; or, have been related to special surface conditions[13] or distribution of charged defects[14], leading to local and sometimes macroscopic noncentrosymmetricity.[2,5,6,11,15]

Recently, the interest in the symmetry breaking in paraelectric phases of ferroelectric materials has been arisen by the discovery of apparently large flexoelectric effect in some ferroelectrics above $T_C$.[16] The flexoelectricity relates polarization and strain gradient[17-19] and appears in all sufficiently insulating materials regardless of symmetry of the material. The high values of the flexoelectric coefficients in paraelectric phases of ferroelectrics cannot be easily reconciled with theoretical predictions.[20,21] It has been suggested[22-24] that the local polar regions interact with strain gradient and/or resulting flexoelectric polarization, contributing to the apparent flexoelectric response.

Alternatively, it has been proposed that macroscopic polarization in a nominally nonpolar state arises from a nonuniform distribution of charged defects in the material.[5] Obviously, there are numerous ways in which this type of symmetry breaking can take place.[6] The best known example are electrets[25] where charge separation is accomplished by applying electric field on the sample and then removing it. One can, however, easily imagine inhomogeneous distribution of charged defects that is not mediated by an electric field.[14,22,26] The resulting polarization could, under certain conditions, contribute to the apparent flexoelectric response. Combination of two or more mechanisms may also be operative in a material. For example, nonuniform charge redistribution may drive preferential orientation of polar regions, as suggested in Ref. 5. Other mechanisms of symmetry breaking of nominally centrosymmetric materials have been demonstrated and discussed in the literature.[2]

Both the wider phenomenon, the breaking of centric symmetry, and ensuing contribution to the pyroelectric, flexoelectric, electrostrictive and/or piezoelectric response of material are of fundamental and possibly also a practical interest. The size of the effects generated by symmetry breaking varies by a few orders of magnitude[22,26] and while the total macroscopic effect is often (but not always[26]) small, large effects may be expected at a small scale, and thus be of interest for nanoelectronics.[27,28] On the fundamental side, understanding mechanisms by which polar regions, nonhomogeneous distribution of charged defects and surface states give rise to symmetry breaking is expected to give a deeper insight into the physics governing their origins.

The general aim of this work is to investigate origins of the symmetry breaking in paraelectric phase of archetypal ferroelectric, $BaTiO_3$.[29] Experimental results indicate that the cubic symmetry of paraelectric $BaTiO_3$ may be broken both microscopically[30] and macroscopically,[8,10,22] however, the underlying mechanisms are not understood. One obvious candidate are polar regions which "condense out" at some high temperature (so-called Burns temperature) and break the cubic symmetry locally.[12] Local dipolar distortions associated with pseudo Jahn-Taller effect have also been proposed as origins of the local symmetry breaking in $BaTiO_3$.[31] To get observed breaking of the macroscopic symmetry, these polar regions (or another entity with a polar symmetry) would have to be aligned on a larger scale by a yet unidentified mechanism.[22] Another candidate are charged defects (oxygen or cation vacancies, aliovalent ionic substitutions) that may be nonuniformly distributed throughout macroscopic sample, giving rise to polarization. In both cases, a mechanism leading to the macroscopic polarization may be related to the strain gradient, as was reported for a related material, $(Ba,Sr)TiO_3$.[22] The study by Biancoli *et al.* [22] also indicated that state of the surface of the sample does not seem to play a dominant role in symmetry breaking in $(Ba,Sr)TiO_3$ solid solution, including $BaTiO_3$ end member. It is important to note here that even though the strain gradient is associated with polarization through flexoelectric effect, one may envision that the strain gradient alone may orient polar regions statically or distribute charged defects, without additional influence of the flexoelectric polarization, which, according to the theory, is expected to be small.[32] One such strain-related mechanism that might contribute to charge redistribution and polarization formation is chemical expansivity.[33-35] A similar interaction with strain gradient can possibly also drive preferential orientation of local polar regions and dipolar pseudo Jahn-Teller distortions. [22,31]

To distinguish between these possibilities, in particular between charged defects and polar-regions models as origin of the macroscopic symmetry breaking, we adopt the following approach. We compare behavior of two polymorphs of $BaTiO_3$, one in which the high temperature phase exhibits a hexagonal structure (h-BTO), with partly face–sharing and partly corner-sharing $TiO_6$ octahedra;[36-38] and the other polymorph with the usual, ideal cubic perovskite structure (p-BTO) with corner–sharing $TiO_6$ octahedra. p-BTO undergoes a sequence of phase transitions from the high temperature centrosymmetric paraelectric-paraelastic cubic phase to a tetragonal phase at $T_C \approx 403$ K, orthorhombic phase at ≈278 K and rhombohedral phase at ≈183 K, all of which are ferroelectric–ferroelastic, with polarization as the order parameter.[39] The indicated temperatures refer to pure material and will be different for cooling and heating runs. h-BTO transforms at $T_{trans} \approx 222$ K from the high temperature centrosymmetric paraelectric–paraelastic hexagonal phase to an orthorhombic noncentrosymmetric (i.e., piezoelectric) paraelectric-ferroelastic phase. h-BTO becomes ferroelectric-ferroelastic below $T_C \approx 60$-74 K,[36,40,41] possibly possessing a monoclinic structure and exhibiting ferroelectric polarization that is about one order of magnitude smaller than in ferroelectric phases of p-BTO.[36,42-46] The order parameter in the ferroelectric-ferroelastic phase of h–BTO appears to be strain.[43] It is thus reasonable to assume that h-BTO should not exhibit polar regions at room temperature or, at very least, their polarization and effect on properties should be much smaller than in paraelectric phase of p-BTO. Assuming that the processing conditions and material parameters such as concentration and nature of charged defects are similar for the two materials, then, absence of properties characteristic for polar structure in h-BTO above Curie temperature and their presence in p-BTO above its $T_C$, would give a support to the models which attribute broken symmetry in p-BTO to the alignment of polar regions. We do not discuss in this paper mechanisms that drive preferential orientation of polar regions or distribution of charged defects and rather use as a starting premise that such mechanisms exist.

For simplicity we will refer to polymorphs of p-BTO and h-BTO by preceding "x-BTO" with terms that characterize the polymorph: for example, "tetragonal p-BTO" designates polymorph of p-BTO with tetragonal structure and "paraelectric-ferroelastic h-BTO" designates polymorph of h-BTO with orthorhombic structure which is ferroelastic but not ferroelectric (the polymorph with orthorhombic crystal structure).

# II. MATERIALS AND EXPERIMENTAL METHODS

## A. Materials preparation and structure

A recently revised version of $BaO-TiO_2$ phase diagram at high temperatures can be found in Ref. 47. p-BTO phase is stable in air up to about ≈1770K, and h-BTO is stable above this temperature up to the melting temperature at ≈1900 K. The low temperature boundary of the stability range of h-BTO depends on partial pressure of oxygen, with lower oxygen pressures stabilizing h-BTO to lower temperatures.[48] Regardless of the sintering atmosphere, h-BTO is metastable below about 1270 K.[48]

We use the same $BaTiO_3$ powder (Inframat Advanced Materials, with particle size of ≈200 nm and purity of 99.95%) to prepare both p- and h-polymorphs of $BaTiO_3$. The X-ray diffraction (XRD) pattern of the as-received powder taken at room temperature reveals that it has crystal structure of the tetragonal p-BTO. The XRD analysis was carried out using a BrukerD8 DISCOVER X-ray diffractometer with monochromatic CuKa1 radiation (1.540596 Å) and a position sensitive detector in the 2q range from 20° to 80°.

The h- and p-BTO samples were prepared using the following two procedures. To prepare h-BTO, disk shaped samples were pressed from as-received powder, which was previously manually mixed with 4% water based solution of polyvinyl alcohol (PVA). The binder to powder ratio was 1:25 in weight. The samples were then heated in air with temperature rate of 5 K/min to ≈1773 K, sintered at this temperature for 4 h, and then let to cool to room temperature with the cooling rate determined by the natural cooling of the furnace. The XRD patterns of those sintered samples taken at room temperature shows a pattern characteristic for hexagonal symmetry of h-BTO (Fig. 1a). No peaks belonging to other phases are apparent. The density of sintered samples was about 92 % of the theoretical. In the second procedure, used to prepare p-BTO, the as–received powder was first milled for 24 hours in isopropanol with $ZrO_2$ balls (5 mm diameter). The powder was then mixed with the same ratio of 4% PVA solution as described above, pressed into pellets, and heated in air with heating rate of 5 K/min to 1723 K. After 4 h sintering the samples were cooled in the furnace to room temperature with the same cooling rate as above. The XRD pattern of those sintered samples measured at room temperature shows a pattern typical for tetragonal phase of p-BTO, Fig. 1b. The density of p-BTO is about 94% of the theoretical. During sintering the samples were placed on $BaTiO_3$ powder and the powder was supported by a Pt foil. Different sintering conditions and environments we tried in our studies[22] have not affected appearance of symmetry breaking and polarization in these and related materials.

For both h- and p-BTO samples the grain size is large, varying from 20 to 50 $\mu$m for p-BTO and 20-80 $\mu$m for h-BTO, with some smaller grains, which is typical for $BaTiO_3$ processed at such high temperatures.[29,49] The reasons for retention of h-BTO phase at room temperature are not known at present and will be discussed elsewhere. The densities of the materials were determined from the measured weight and volume.

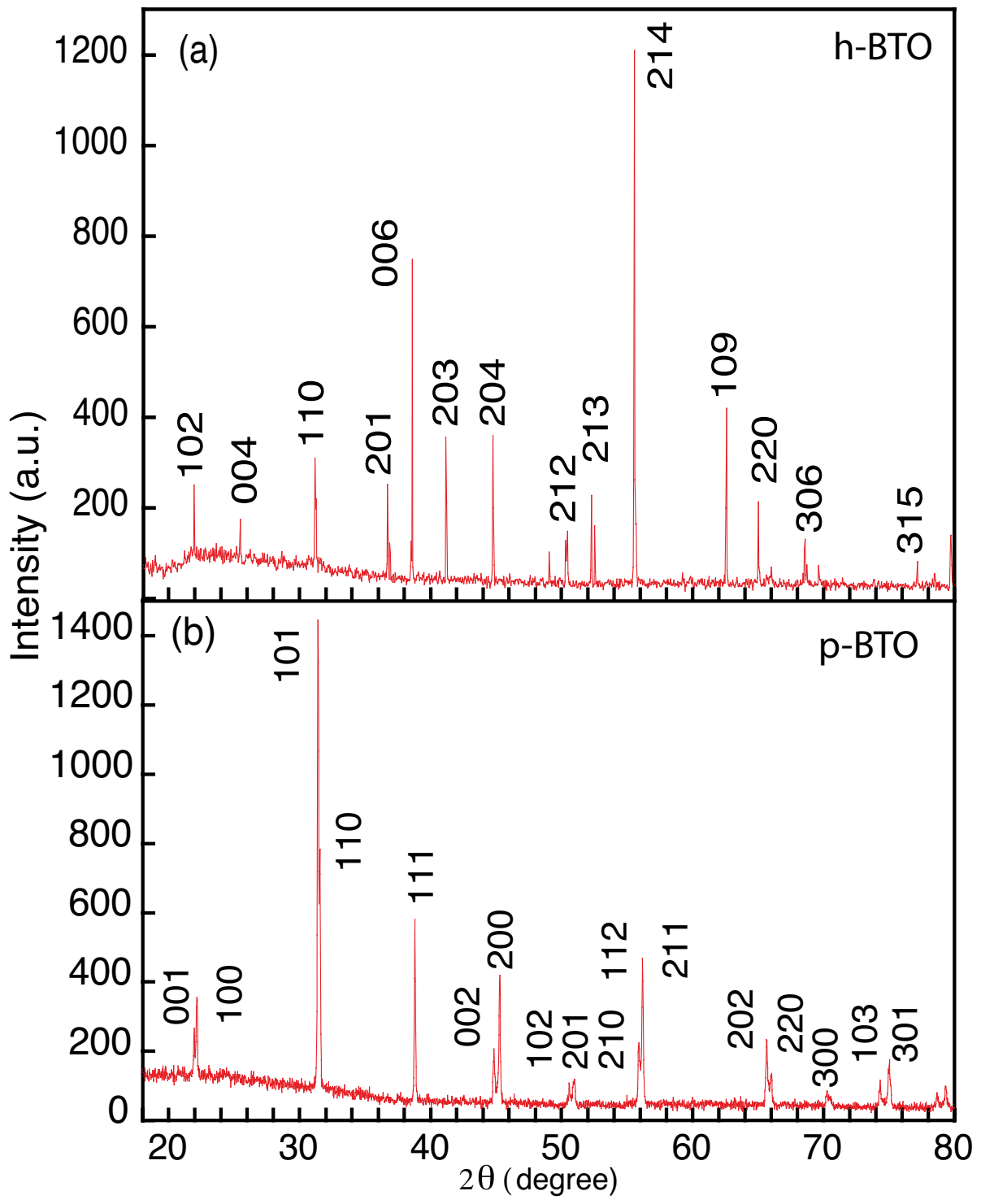

Figure 1. XRD patterns for a) h-BTO and b) p-BTO. Indices for some peaks are not shown. The peaks were indexed using JCPDS cards 34-0129 for h-BTO and 5-0626 for p-BTO.

## B. Measurement methods for electrical and mechanical properties

For electrical measurements, Au electrodes were sputtered on polished major surfaces of the disk shaped samples. The thickness of the samples ranged from 0.4 to 0.5 mm. The dielectric permittivity was measured as a function of temperature at several frequencies of the driving field with amplitude 1 $V_{rms}$ using an HP 4284A Precision LCR meter while sample was placed in a Delta 9023 temperature chamber or a CTI-cryogenics refrigerator model 22. Measurements were made during cooling, with the rate of 2 K/min or less. Dielectric nonlinearity was investigated by measuring capacitive current of a sample, as a function of the amplitude of driving voltage, $V = V_D \sin(\omega t)$. The current was determined by measuring voltage $V_S$ on a standard resistor R placed in series with the sample. The permittivity was determined from the capacitance which is in turn given by $C = V_S / (V_D R\omega)$. R was chosen such that $R << 1/\omega C$. $V_S$ was measured with a SR830 Lock-in Amplifier. Typical frequencies were 1 kHz.[50]

The mechanical stiffness (storage modulus and loss) was measured in the single cantilever mode with a Perkin-Elmer PYRIS Diamond dynamic mechanical analyzer, during cooling with the rate of 1-2 K/min. The pyroelectric current was measured using custom-made set-up and dynamic method described in detail in Ref. 51. The current was excited by cycling temperature of the sample with a Peltier element. A typical temperature waveform was triangular with temperature amplitude of 0.5 to 1 K and frequency of 20 mHz.

Thermally simulated currents (TSC) were measured by collecting current while sample was heated on a top of a custom-made hot plate from room temperature to 823 K with at constant heating rate of 2.5 K/min, followed by cooling at the same rate. No electric field was applied on samples either before or during TSC measurements. The current was measured using a Keithley 486 picoammeter. The samples are placed on Pt foil, which was insulated electrically from the hot plate with a sapphire wafer. Pt foil was grounded and the current collected from the top electrode of the sample.

## III. RESULTS AND DISCUSSION

### A. Dielectric permittivity as a function of temperature

Figures 2a and 2b compare relative unclamped permittivities of p-BTO and h-BTO samples. The temperature dependences of the measured permittivities are similar to those reported by other authors,[36,44] confirming the XRD data that the two kinds of materials are indeed p- and h-BTO. The peaks in permittivities correspond to structural phase transitions in these two materials. It is important for the later discussion to notice much lower permittivity in h-BTO compared to that in p-BTO. The dielectric loss factor tan$\delta$ (not shown) at room temperature and at 1 kHz is 0.01 in p-BTO and 0.001 in h-BTO. The higher loss in p-BTO is at least in part related to the presence of ferroelectric domain walls.

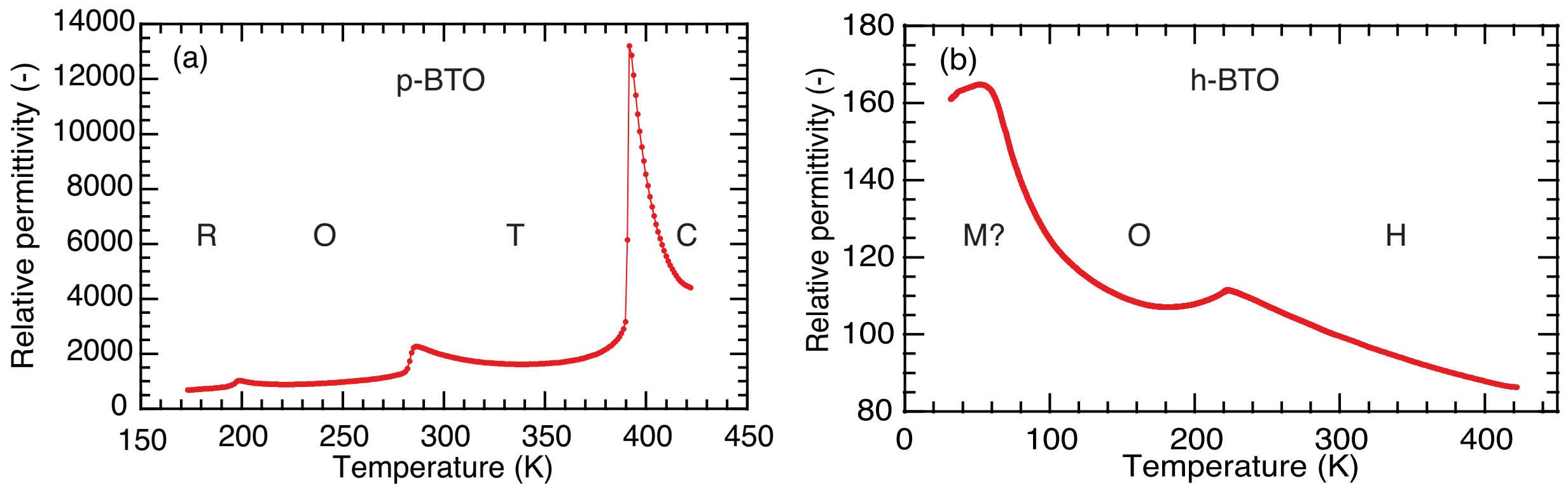

Figure 2. Temperature dependence of the relative permittivity of a) p-BTO and b) h-BTO. Letters R, O, T, C, M and H stand, respectively, for phases with rhombohedral, orthorhombic, tetragonal, cubic, monoclinic and hexagonal symmetries. Our measurements indicate that maximal permittivity in h-BTO is reached between 50 and 60 K, in agreement with Refs. [40,41], while Ref. 36 reports $T_C \approx 74$ K. The discrepancy could be related to the purity of the materials or accuracy of temperature measurements at cryogenic temperatures.

## B. Elastic properties as a function of temperature

Elastic properties of p-BTO have been previously studied for signs of polar precursors within cubic phase using Resonance Piezoelectric Spectroscopy.[52] Presence of dynamic polar regions above the Curie temperature has been deduced by observing resonances which the authors associated with piezoelectricity of the polar regions. This signature of polar regions was detected at temperatures as high as 613 K. Polar regions in incipient ferroelectric-ferroelastic materials, such $Pb(Mg_{1/3}Nb_{2/3})O_3$, [53] are both electrically and elastically active, i.e., they possess both polarization and strain with respect to the matrix of the parent phase.[54] In analogy to polar precursors of a ferroelectric phase, one may therefore expect to see[55] non-polar precursors of the ferroelastic phase in h-BTO above ≈220 K. Figure 3 shows elastic modulus and associated tan$\delta$ as a function of temperature for h-BTO samples. Transition from the hexagonal paraelectric-paraelastic phase into orthorhombic paraelectric-ferroelastic phase is clearly visible around 220 K. The transition is characterized by a large softening[56] of the material possibly related to formation of ferroelastic domains below the phase transition temperature. More interesting is the frequency dispersion of the elastic modulus and frequency dependent peak of mechanical tan$\delta$ above the phase transition temperature. Both are reminiscent of effect of polar regions on permittivity and elastic modulus[54] in relaxor ferroelectrics (note that we are here plotting stiffness, and not susceptibility (compliance)).

The elastic data, therefore, suggest formation of nonpolar precursors of ferroelastic phase in h-BTO above ≈220 K. It is important to recall that ferroelastic phase of h-BTO is noncentrosymmetric i.e., piezoelectric. This means that local nonpolar precursors of the ferroelastic phase may exhibit piezoelectric but not pyroelectric effect. Consequences of this are discussed in more detail in the next sections.

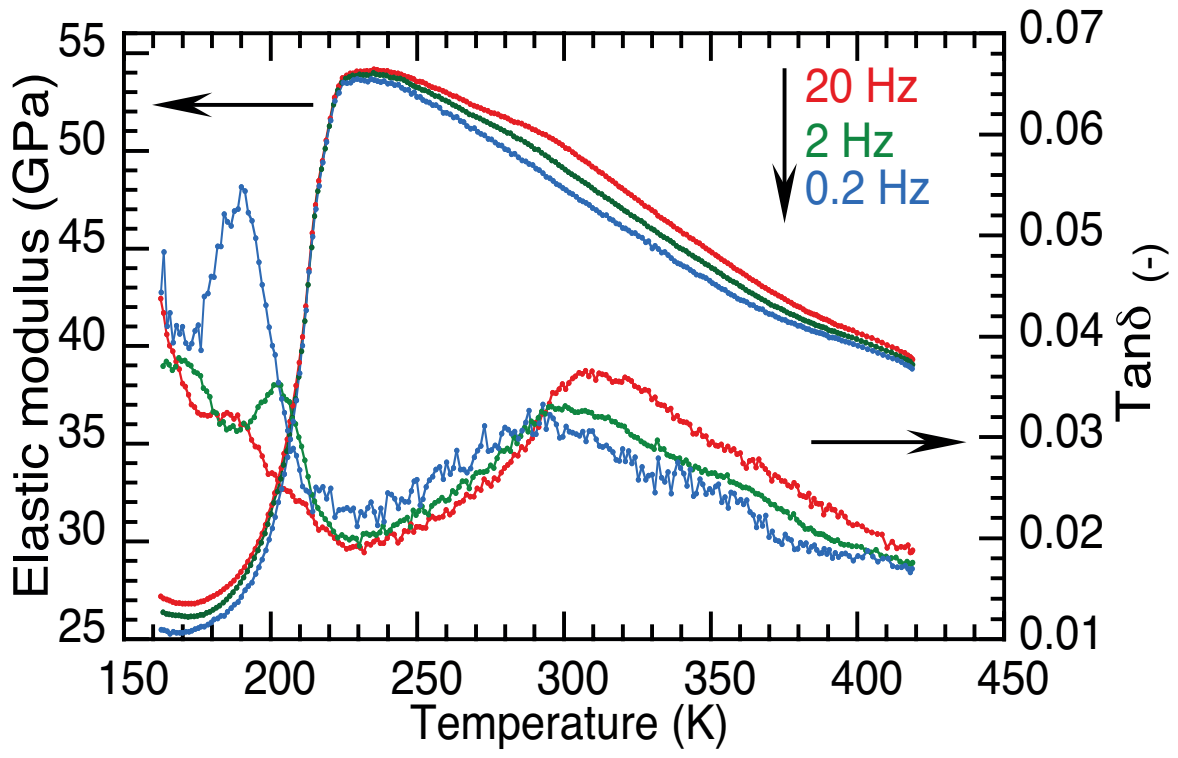

Figure 3. Elastic modulus and loss factor for h-BTO. The data for modulus are in a good agreement with those reported in Ref. 40.

## C. Dielectric nonlinearity

It is known that polar nano regions in relaxor ferroelectric materials[57] contribute to dielectric nonlinearity and show dynamics similar to that observed by domain walls in ferroelectrics.[58-60] The dielectric nonlinearity, if of the right type, may thus be used as an indirect evidence of presence of polar regions in the paraelectric phase of a ferroelectric. In their study of $(Ba,Sr)TiO_3$ thin films, Garten and Trolier-McKinstry[24,61], showed that above $T_C$ the dielectric permittivity of their films exhibits dependence on electric field that is similar to the one observed in the ferroelectric phase. They propose, therefore, that the nonlinearity is due to motion of polar regions and that this motion can account for the apparently large flexoelectric effect reported for this material. A similar mechanism for the large flexoelectricity is also proposed in Ref. 23.

It is of interest therefore to investigate whether paraelectric phases of p-BTO and h-BTO exhibit dielectric nonlinearity; the result could be used to infer presence or absence of polar precursors. If the dielectric nonlinearity is caused by dynamics of polar regions, one would not expect to see it in h-BTO but could be apparent in p-BTO. This is not a trivial point. Polar regions in chemically simpler compounds, such as $BaTiO_3$, probably do not have the same properties as in more complex perovskites with mixed cations, such as $(Ba, Sr)TiO_3$ and $Pb(Mg_{1/3},Nb_{2/3})O_3$, and could manifest themselves in a less obvious way.

Figure 4 shows normalized dielectric permittivity for p-BTO, at two temperatures above $T_C$, and for h-BTO, at room temperature and just below the temperature of the phase transition into ferroelastic orthorhombic phase (both temperatures are above $T_C$ for h-BTO). The nonlinearity in p-BTO, Fig. 4a, appears to be very small, but this is expected at relatively weak fields used in the experiment. In fact, for a similar field range, the nonlinearity in p-BTO is not much different than that reported in Ref. 61 for $(Ba,Sr)TiO_3$. The nonlinearity in p-BTO is lower at 443 K than at 413 K (closer to $T_C$), which is consistent with a general expectation that polar regions should be larger and more numerous closer to Curie temperature. The origin of nonmonotonous field dependence of the permittivity for p-BTO has not been investigated.

An even smaller, but measurable nonlinearity has also been observed in h-BTO, Fig. 4b. It is tempting to explain the small nonlinearity at room temperature for this material as being associated with precursors of ferroelastic state, which, while not polar would be piezoelectric. Those nonpolar precursors could therefore contribute to the apparent permittivity through electro-mechanical coupling. However, this conjecture does not seem to be supported by the data taken at 200 K: if the piezoelectricity of ferroelastic domains and precursors contributes to the dielectric nonlinearity, the nonlinearity should be stronger at 200 K, just below the phase transition temperature into ferroelastic/piezoelectric orthorhombic phase, than at room temperature, while the opposite is observed. One could argue that ferroelastic domain walls may be partially frozen at 200 K and not being able to move under weak fields used in the study.[62] One cannot exclude other origins of the weak dielectric nonlinearity in h-BTO, such as small amounts of residual p-$BTiO_3$ phase.

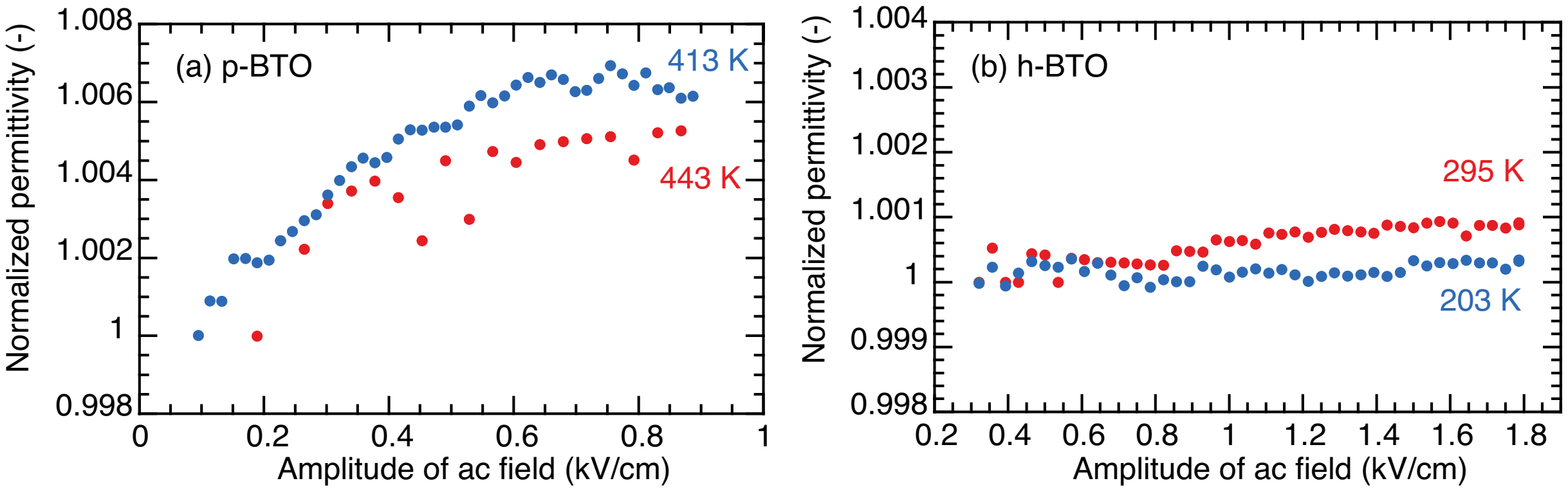

Figure 4. The dependence of dielectric permittivity on amplitude of the electric field for a) p-BTO and b) h-BTO. The values of permittivity are normalized with respect to the permittivity measured at 10 V. All voltages are in rms.

## D. Pyroelectric properties

The elastic[52] and optical[30] data from the literature and those presented in the previous section indicate presence of precursors of ferroelectric state within the cubic phase of p-BTO and non-polar precursors of ferroelastic state within hexagonal phase of h-BTO. These precursors of low-temperature ferroic phases break locally centric symmetry of each parent phase.

We now verify whether the precursors of ferroic states break macroscopic symmetry of cubic p-BTO and hexagonal h-BTO. Two statements can be made *a priori*: The nominally cubic symmetry of p-BTO may be broken macroscopically if polar regions are preferentially statically oriented throughout the sample. An aggregate of ordered polar regions could exhibit nonzero macroscopic polarization, albeit very small. This polarization may be detected in a nondestructive way by measuring pyroelectric effect i.e., change of the polarization with respect to a small change in temperature. In the case of hexagonal h-BTO, the samples are not expected to exhibit macroscopic pyroelectric effect because ferroelastic precursors in h-BTO are piezoelectric, but not polar. Thus, even if precursors are preferentially oriented in hexagonal phase of h-BTO one should not observe macroscopic pyroelectricity (we neglect here secondary and tertiary pyroelectricity[63]).

The pyroelectric response of cubic p-BTO measured at ≈413 K is shown in Fig. 5a ($T_C$ during cooling is 391 K, Fig. 2a). Clear modulation of the current with temperature can be observed. Drift of the pyroelectric current with time and noise in the current are due to fluctuations in the background current and instability of the background temperature. The same experiment is shown for h-BTO at ≈294 K in Fig. 5b. h-BTO sample does not exhibit pyroelectric current, even if the temperature amplitude is two times higher than used for p-BTO. These experimental results are thus consistent with the picture of ordered polar precursor present in the cubic phase of p-BTO and absent in the hexagonal phase of h-BTO. We note that the pyroelectric current is also observed in single crystals of p-BTO[22], indicating that it is not due to polycrystalline nature of the samples examined in the present work.

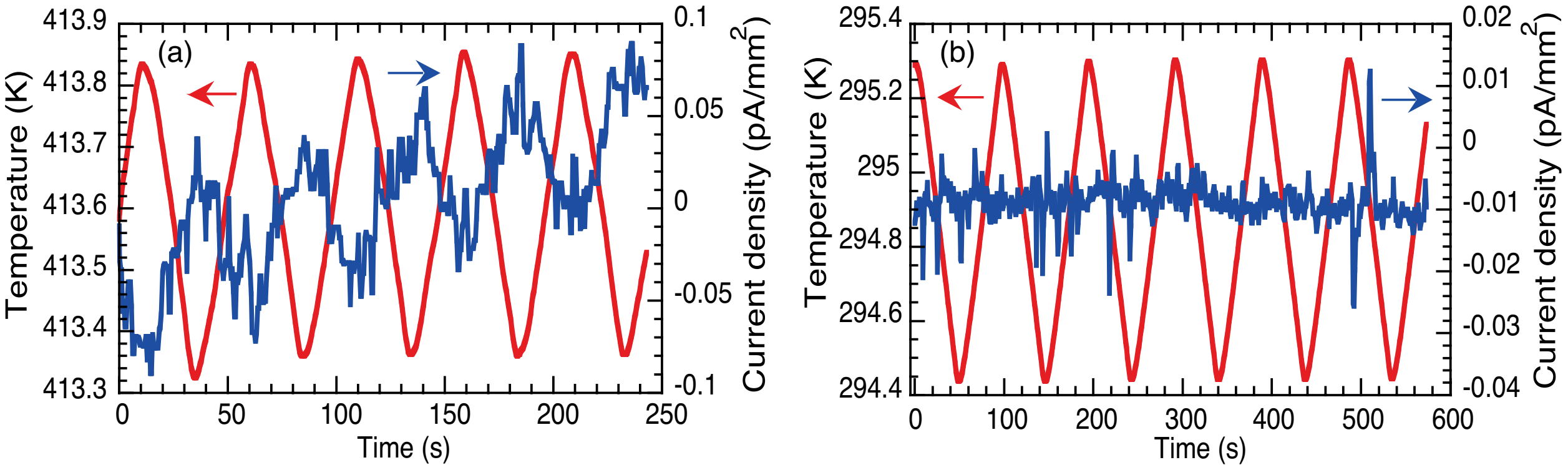

Figure 5. The pyroelectric current and the triangular temperature waveform for a) p-BTO at 413 K (20 K above $T_C$) and b) h-BTO at room temperature (230 K above $T_C$).

These results are also in agreement with other reports in the literature, where various effects forbidden by the nominal cubic symmetry were demonstrated in p-BTO indicating global symmetry breaking.[8,52] The situation in h-BTO is more complex and it is possible to tentatively advance several arguments, in addition to absence of polar regions, why pyroelectric response is not seen in this material. We will address those issues in Section F.

## E. Macroscopic symmetry breaking revealed by Thermally Stimulated Currents

A possible evidence of macroscopic ordering of local polar precursors in cubic phase of p-BTO has been demonstrated by detecting pyroelectric response in this material (Section D). The pyroelectric data thus suggest that both local and global centric symmetry of cubic p-BTO are broken. The absence of pyroelectric current in hexagonal phase of h-BTO samples is in agreement with absence of polar precursor, as is expected considering the symmetry of that material. However, pyroelectric data do not give information whether macroscopic centric symmetry is broken in h-BTO; those data only suggest that material is not macroscopically pyroelectric (polar).
To look for signatures of macroscopic breaking of centric symmetry that does not manifest itself in the pyroelectric effect, we measured TSC. Samples of p-BTO and h-BTO were cut into halves. Surfaces of the samples were marked with respect to their position in the furnace in which samples were sintered:[22] one side of the disk was marked as "up" and the opposite side as "down". During TSC measurements, the "up" side of one half of a sample was placed on a Pt foil, which was grounded. In the next experiment, the other half of the sample was placed with "down" side on the Pt foil. The two halves of each sample were, thus, flipped by 180° with respect to electrodes. The electric current was measured in the samples while the temperature was increased and then decreased. Both materials exhibited thermally stimulated currents, as shown in Fig. 6 during heating and in Fig. 7 during cooling. Since no electric field was applied on samples during the measurements and samples were not polarized by an electric field before measurements to form an electret, the question can be posed about origins of the currents and nature of the peaks in the currents.

p-BTO is pyroelectric and the current and peaks seen in TSC around 550-650 K, Fig. 6, may be due to depolarization of the sample. However, no anomaly in current is observed during cooling, Fig. 7, and samples are still pyroelectric after cooling to room temperature, as verified by pyroelectric measurements (not shown). If the current is due to depolarization and polarization is still present in the sample after being cooled, the peaks in current should appear also during cooling (with opposite sign). Another possibility is that the currents originate from the thermopolarization effect,[64,65] which may appear in samples with inversion center and could thus be observed in both p- and h-BTO. To be observed, however, the themopolarization current requires a large, time–dependent temperature gradient. In addition to thermopolarization current, the current may originate from the small voltage burden on the picoammeter (about 300 $\mu$V in case of device used in our experiments) or from a temperature gradient across the sample (potential difference of Seebeck effect).[66] Let us assume that the current originates from one of these three sources. For a sample with a perfect centric symmetry, the two orientations of a sample, "up" and "down", should be equivalent and the response to either the voltage or the temperature gradient should be the same, regardless of the orientation of the sample. Therefore, the direction of the currents in the halves of each sample should be the same. What is observed here and in other related materials[22] is that direction of the current is sensitive to orientation of the samples (see Fig. 6, the region from 550 to 650 K). After eliminating obvious possibilities such as surface effects, asymmetrical electrodes and like[22], this experiment proves that the samples cannot possess macroscopic centric symmetry. While the asymmetry in TSC may be expected for p-BTO because it exhibits macroscopic pyroelectricity and is therefore asymmetric, the sensitivity of TSC on orientation of h-BTO samples is surprising.  These experiments shows unequivocally that macroscopic inversion symmetry is also broken in h-BTO samples.

It is interesting to mention here that in their seminal work[67] Bucci *et al.* reported that some Teflon samples exhibited temperature dependent currents in nonpolarized samples and that currents changed sign when the samples were inverted with respect to electrodes. In other aspects, though, Teflon behaved differently than samples investigated here.

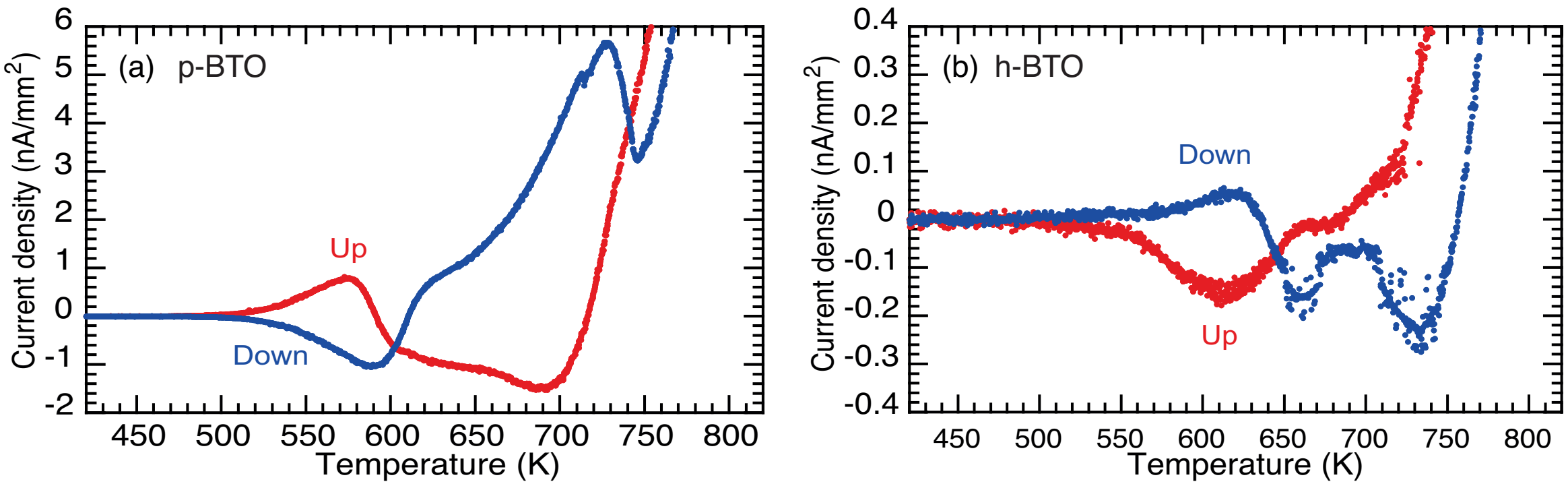

Figure 6. TSC for a) p-BTO and b) h-BTO. Note the difference in the sign of the current peaks in the region 550-650 K, indicating that direction of the current depends on the orientation of the sample with respect to the electrodes. See Ref. 22 for more details.

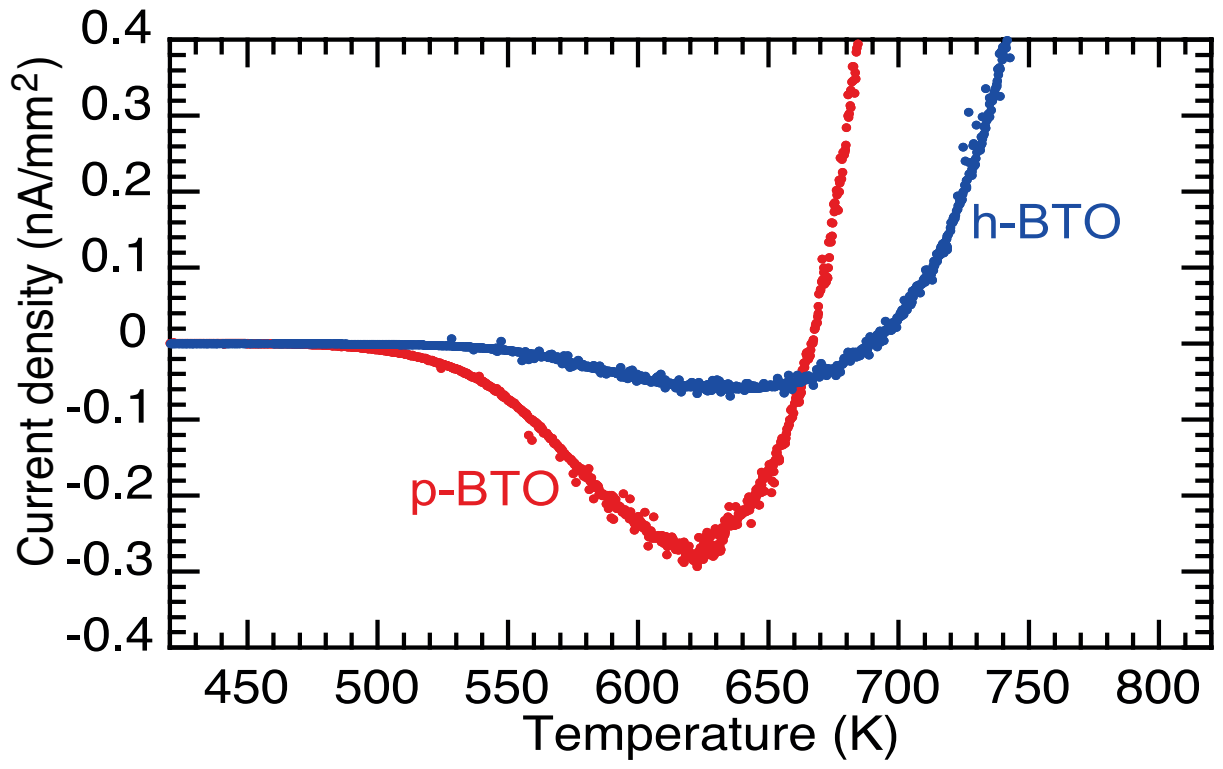

Figure 7. TSC for h-BTO and p-BTO during cooling, In both samples the "down" side is grounded and faces the hot plate.

## F. Discussion

The primary purpose of our study is to investigate contributions to the processes that lead to appearance of macroscopic polarization in centrosymmetric phases of examined materials; specifically, we aim to distinguish between polar regions, on one side, and other mechanisms, such as inhomogeneous distribution of charged defects and built-in flexoelectric polarization, on the other side. We start this Section by summarizing the evidence discussed and presented in the previous Sections. The present combined study of dielectric, elastic, and pyroelectric properties of p-BTO and h-BTO, together with the knowledge of the crystal symmetry of the two materials and previous studies on polar regions in p-BTO, indicate that: (i) cubic phase of p-BTO exhibits local and macroscopic breaking of nominal centric symmetry and exhibits local and global polarization. Both are likely linked to polar regions, precursors of the ferroelectric phase; (ii) hexagonal phase of h-BTO does not exhibit local polar regions and macroscopic polarization, although its macroscopic centric symmetry is broken. h-BTO probably exhibits precursors of the nonpolar paraelectric-ferroelastic phase; (iii) precursors of the polar phase are mobile in the paraelectric phase of p-BTO, while precursors of paraelectric-ferroelastic phase are elastically active in hexagonal h-BTO. The polar regions in p-BTO appear to be macroscopically ordered, resulting in the global polarization of this material.

The mechanism for alignment of polar regions in p-BTO can be sought in a strain gradient, similar to the one that has been measured in related $(Ba,Sr)TiO_3$ solid solution, and which is induced in the samples during the densification at high temperatures (or during growth, in case of single crystals).[22] Charged defect gradient may also drive orientation of polar regions, as proposed[5] to explain self–polarization in relaxor $Pb(Mg_{1/3}Nb_{2/3})O_3$. Since p-BTO and h-BTO are prepared under similar conditions, from the same powder, the strain gradient may be expected also in h-BTO, but ordering of paraelectric-ferroelastic phase precursors does not result in macroscopic polarization because those precursors are not polar. The strain (or another type of) gradient is, however, needed in the qualitative model of both materials because it explains straightforwardly the breaking of inversion symmetry that is manifested in h-BTO through TSC. A charge gradient is not expected in h-BTO because it would itself lead to polarization, which is not observed in this material.

One may further discuss a possibility that, because h-BTO is ferroelectric below $T_C$≈52 K-74 K (See Refs. 36,40,41 and Fig. 2), the precursors of the paraelectric-ferroelastic phase could carry polar distortion, anticipating the low-temperature ferroelectric phase. Such effect may not be easily seen in pyroelectric measurements because room temperature is too far away from the $T_C$ of h-BTO and density of polar precursors should scale with distance $T-T_C$ from the transition temperature. This is an interesting possibility and atomic resolution techniques should be employed to answer a question on polarity of ferroelastic regions in hexagonal h-BTO. We do repeat (see Introduction), however, that the polarization is very weak in the ferroelectric phase of h-BTO, and the order parameter is strain rather than polarization.[43] Thus it is unlikely that ferroelastic regions would exhibit a significant polarization that could not be seen in our pyroelectric measurements. As an indication of the detection limit of pyroelectric current that might originate from ordered polar regions, we measured at room temperature a rather strong pyroelectric current in polycrystalline $(Ba_{0.1}Sr_{0.9})TiO_3$ with $T_C$ ≈65 K, i.e. in the same region as in h-BTO, some 230 K above its Curie temperature.[68] On the other hand, in $(Ba_{0.025}Sr_{0.975})TiO_3$ with $T_C$ ≈25 K[69] and which, like pure $SrTiO_3$, is not likely to exhibit polar regions, we did not see pyroelectric effect at room temperature. All experimental data presented can thus be consistently interpreted in terms of the strain gradient in the samples and its interaction with microscopic precursors of the ferroic states of the two materials. Verification of this conjecture is now subject of further studies.

We next consider alternative mechanisms for the symmetry breaking. It should be recognized that a strain gradient itself breaks the symmetry of a sample with centric symmetry. Even if polar regions are not available, the origin of the pyroelectric effect could be built-in flexoelectric polarization, which by definition accompanies the strain gradient. A similar mechanism was suggested for thin films, due to either chemical or strain gradient.[70] In such a case, the absence of pyroelectric current in h-BTO implies that pyroelectric signal associated with built-in flexoelectric polarization could be too small to be measured. That, in turn could be related to the small permittivity of h-BTO, because it is known that flexoelectric polarization is proportional to the intrinsic permittivity of the material.[19] We propose the following counterargument: as already mentioned above, a rather large pyroelectric response was measured in $(Ba_{0.1}Sr_{0.9})TiO_3$, whose relative permittivity at room temperature (≈300 at 1 kHz) is not much higher than that in h-BTO (≈100). On the other hand, $(Ba_{0.025}Sr_{0.975})TiO_3$ with nearly the same relative permittivity (≈260 [68] ) as $(Ba_{0.1}Sr_{0.9})TiO_3$ does not exhibit pyroelectric current at room temperature. The difference between $(Ba_{0.025}Sr_{0.975})TiO_3$ and $(Ba_{0.1}Sr_{0.9})TiO_3$ could be that the former, like $SrTiO_3$, does not possess polar regions that could enhance its apparent flexoelectric response.[23,71] Analogously, the absence of pyroelectric current in h-BTO thus rather suggests absence of a particular contributor (i.e., polar regions) rather than insufficient resolution of measurements. Interestingly, presence of built-in polarization was ruled out in some flexoelectric experiments on single crystals of $BaTiO_3$.[72]

It can be argued that the macroscopic polarization is due to separation of charged cationic and anionic vacancies or to inhomogeneous distribution of charged defects, rather than alignment of polar regions.[68] Separation of defects could be driven by the strain gradient via chemical expansivity effect[22,33-35] while, on the other hand, the strain gradient could facilitate creation of ionic vacancies[35]. The origin of vacancies could be in precipitation of small amounts of secondary phases, such as barium orthotitante, as reported in Ref. 22. The peaks in TSC observed in the 550-650 K range, which differ in intensity by an order of magnitude between p-BTO and h-BTO, Figure 6, and dielectric loss data (Section III.A) suggest that h-BTO possesses a lower concentration of mobile charges than p-BTO, even though two materials were prepared from the same powder and at similar sintering conditions. This is somewhat surprising considering that h-BTO is usually stabilized under reducing conditions and one would expect a higher concentration of oxygen vacancies and reduced Ti in h-BTO.[48,73] Furthermore, the higher temperature at which TSC peaks appear in h-BTO than in p-BTO, Figure 6, shows that trapped charges whose release is responsible for TSC peaks, have a higher activation energy in h-BTO than in p-BTO.[25] This could be related to different connectivities of oxygen octahedra in the two materials. All this, together with a much lower permittivity of h-BTO, may make associated polarization too rigid and too small to respond to the weak temperature signal during pyroelectric measurements at room temperature. If so, one could explain different behavior of p- and h-BTO without invoking polar regions. We believe that this possibility should be further investigated. Concerning a possible link between the low permittivity and the charge model, it is interesting to mention that we have measured a clear pyroelectric response in polycrystalline As-doped $Pb_3(PO_4)_2$ (not shown). Like h-BTO, this material is a nonpolar ferroelastic and it exhibits at room temperature a relative permittivity on the order of 20-50[74], which is even smaller than in h-BTO. It appears that a low permittivity alone cannot explain why pyroelectric effect was not observed in h-BTO.

We address briefly the origin of the peaks in TSC (Fig. 6), which we have used in this study only to demonstrate absence of inversion center in the samples. As already stated in Section III. E, the peaks in current are not due to depolarization of the samples. This conclusion can be made because peaks are not observed during cooling (Fig. 7) and the pyroelectric signal in p-BTO and other materials can be observed immediately after cooling samples to room temperature; that is, polarization, whatever its origin, is not lost during heating of the samples to 823 K. We have strong evidence, as discussed in detail in Refs. 22,68, that peaks are due to trapped charges whose activation energy is rather high, and which need time to resettle into traps during cooling. These charges may reinforce built-in polarization but are not its origin. The direction of the thermally stimulated current originating from the charge de-trapping is sensitive to sample orientation probably due to the strain gradient[75,76], but this mechanism is not yet understood.

Another conclusion that can be drawn from the TSC experiments and absence of polar regions in h-BTO, is that the peaks in the currents are most likely not related to polar regions of p-BTO and precursors of ferroelastic state in h-BTO. We refer here to experiments in relaxor ferroelectrics, where a peak in thermally activated acoustic emission is seen at one or more characteristic temperatures below Burns temperature and has been associated with processes within polar nano regions.[77,78] One could speculate that the peaks in TSC observed here correspond to those acoustic emission peaks and have the same origin. Since the acoustic emission peaks would be generated from both polar regions and precursors of ferroelastic state, one would expect to see acoustic emission peaks and peaks in TSC in both p- and h-BTO. In h-BTO, in which ferroelastic phase is piezoelectric but not polar, the electric charges needed for the current peaks could originate from the piezoelectric effect. In experiments on relaxor ferroelectrics, however, acoustic emission peaks were observed during cooling[77] while we do not see any peaks in TSC during cooling, Fig. 7. Thus, the two phenomena (TSC and acoustic emission peaks) do not appear to be closely linked.

We finally briefly comment on possibility that the polarization in nominally centrosymmetric BTO may be due to pseudo Jahn-Teller effect (PJTE) where local dipolar distortions interact with the strain gradient, as suggested in Ref. 31. Since we do not see polarization in h-BTO, that could indicate differences in the PJTE in hexagonal h-BTO and cubic p-BTO. Our experimental results could thus be a test for theoretical models of PJTE in hexagonal BTO and for the role of PJTE in enhancing properties of ferroelectric-based materials suggested in Ref. 79.

**Acknowledgement:**

This work is supported by the Swiss National Science Foundation (No. 200021-159603). Samples of As-doped $Pb_3(PO_4)_2$ were kindly supplied by E.K.H. Salje.